\documentclass[smallextended]{svjour3}       
\smartqed  
\usepackage{amsmath}
\usepackage{amssymb}
\usepackage{graphicx}
\usepackage[]{hyperref}
\hypersetup{
    pdftitle={necklace},
    pdfauthor={A. S. Bradley},
    pdfsubject={necklace},
    colorlinks=true,
    linkcolor=blue,
    citecolor=blue,
    filecolor=black,
    urlcolor=blue
}
\def\urlprefix{}
\def\url#1{}
\usepackage{txfonts}
\usepackage{soul}
\usepackage{color}
\def\g2d{\varg_{\rm 2D}}
\newcommand{\for}{\textrm{for}}
\newcommand{\eqa}[1]{\begin{eqnarray}#1\end{eqnarray}}
\newcommand {\eref}[1]{(\ref{#1})}

\begin{document}

\title{Neutral vortex necklace in a trapped planar superfluid}

\author{M. M. Cawte \and
        M. T. Reeves \and A. S. Bradley
}


\institute{M. M. Cawte \and A. S. Bradley \at
Dodd-Walls Centre for Photonic and Quantum Technologies, Department of Physics, University of Otago, Dunedin 9016, New Zealand. \\
           M. T. Reeves \at
           ARC Centre of Excellence in Future Low-Energy Electronics Technologies (FLEET), School of Mathematics and Physics, University of Queensland, St. Lucia, QLD 4072, Australia.
}

\date{Received: date / Accepted: date}

\maketitle

\begin{abstract}
  We study quantum vortex states consisting of a ring of vortices with alternating sign, in a homogeneous superfluid confined to a circular domain. We find an exact stationary solution of the point vortex model for the neutral vortex necklace. We investigate the stability of the necklace state within both the point-vortex model and the Gross-Pitaevskii equation describing a trapped atomic Bose-Einstein condensate at low temperature. The point-vortex stationary states are found to also be stationary states of the Gross-Pitaevskii equation provided the finite thickness of the outer fluid boundary is accounted for. Under significant perturbation, the  Gross-Pitaevskii evolution and point-vortex model exhibit instability as expected for metastable states. The perturbed vortex necklace exhibits  sensitivity to the perturbation, suggesting a route to seeding vortex chaos or quantum turbulence.
\keywords{Vortex \and Bose-Einstein condensate\and Superfluid}
\end{abstract}

\section{Introduction}
\label{intro}
The point-vortex model (PVM) of planar fluid motion~\cite{helmholtz_lxiii._1867,kirchhoff_gustav_robert_vorlesungenuber_1876} provides a fascinating window into fluid dynamics: it offers an approximate model of normal fluid turbulence~\cite{novikov_dynamics_1975}, and finds a rigorous manifestation in planar superfluids~\cite{onsager_statistical_1949,gauthier_giant_2019,johnstone_evolution_2019} where vorticity is quantized to integer multiples of $h/m$, where $h$ is Planck's constant and $m$ is the atomic mass. Dilute gas Bose-Einstein condensate (BEC) provide a favorable system in which to realize PVM physics. Precision confinement, forcing, and imaging of vortices in BECs have enabled experimental studies of vortex dipoles~\cite{neely_observation_2010}, the von Karman vortex street~\cite{sasaki_benardvon_2010}, and observations of Onsager's negative-temperature vortex cluster states~\cite{gauthier_giant_2019,johnstone_evolution_2019}. The PVM regime can be approximately realized in an atomic BEC, despite the compressible nature of the BEC superfluid~\cite{pismen_vortices_1999,bradley_energy_2012}. Advances in spatiotemporal control~\cite{henderson_experimental_2009,gauthier_direct_2016} allow disc-shaped hard-wall traps to be created with dynamic laser manipulation~\cite{samson_deterministic_2016}. Such advances offer exquisite control of the creation and placement of quantum vortices, and may soon enable systematic experimental study of dynamics and statistical mechanics in quasi-homogeneous systems~\cite{aioi_controlled_2011,reeves_signatures_2014,simula_emergence_2014,yu_theory_2016,yu_emergent_2017,cawte_snells_2019,groszek_decaying_2020}, or forcing tailored to drive an inverse-energy cascade~\cite{bradley_energy_2012,reeves_inverse_2013}. 

A number of interesting links between stationary point-vortex systems in the infinite plane and the roots of special orthogonal polynomials \cite{aref_vortices_2007} have been identified, including to Hermite \cite{stieltjes_sur_1885}, Laguerre \cite{aref_equilibrium_1995}, and Adler–Moser \cite{adler_class_1978} polynomials. Yet the stationary and periodic solutions in the disc have received less attention. In this work, we find an exact analytical solution of the PVM describing a class of neutral vortex necklace states in a disc-confined atomic BEC. The neutral vortex necklace is determined by the roots of a quadratic polynomial related to the \textit{metallic means}, a generalization of the golden ratio. We compare the exact PVM solutions with dynamical simulations of the Gross-Pitaevskii equation (GPE), finding that the stationary states in the PVM are also stationary GPE solutions, provided a small correction is applied to allow for the finite thickness of the fluid boundary. We also carry our further simulations of the GPE for weakly and highly deformed vortex necklace states and compare with simulations in the PVM. 

The structure of the paper is as follows. We first review the PVM in Section \ref{sec:pvmodel}. In Section \ref{sec:metalic_necklace} we derive the stationary solution for the neutral vortex necklace. In Section \ref{sec:simulation} we present our simulations of the vortex necklace and its perturbations, comparing the PVM and GPE results. In Section \ref{sec:conclusions} we conclude.

\begin{figure}[t!]
  \centering
    \includegraphics[width=0.7\columnwidth]{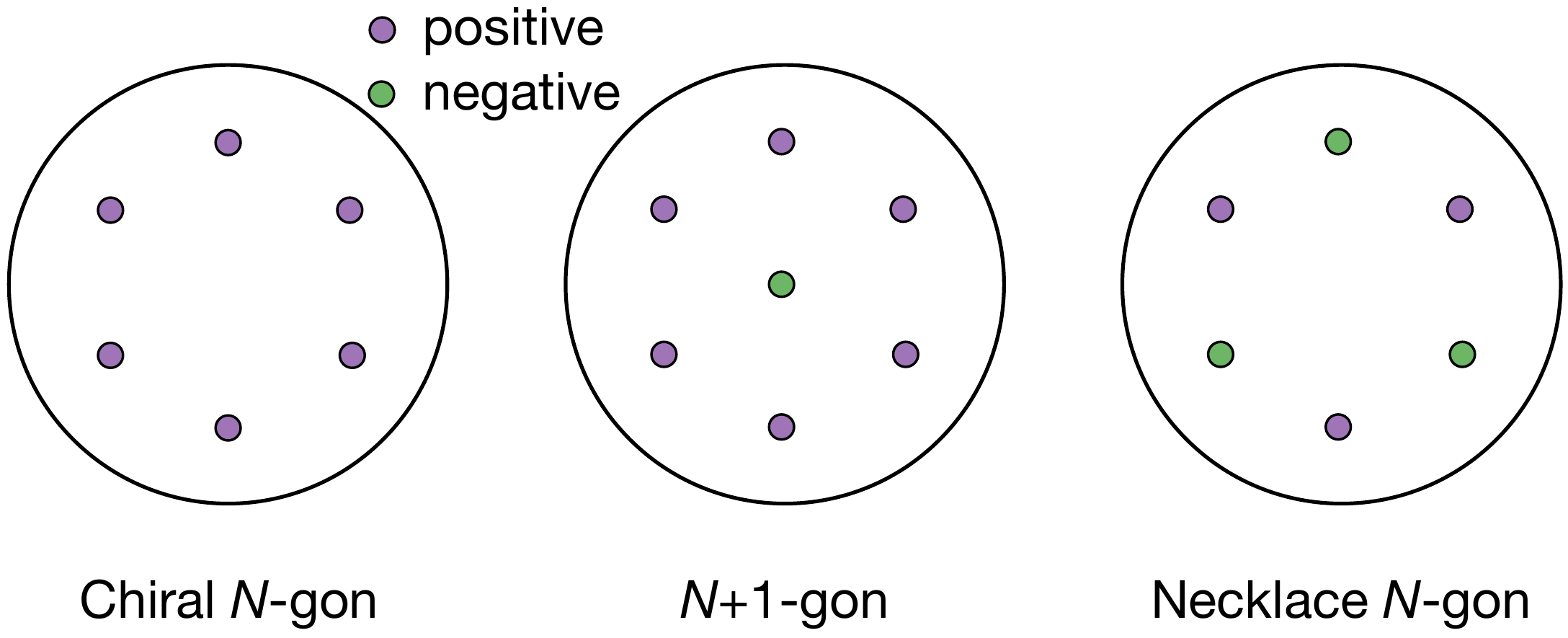}
    \caption{Schematic of vortex $N$-gon, $N+1$-gon, and Necklace $N$-gon states confined to a disc, for $N=6$. The chiral state rotates rigidly without a change in shape, and the $N+1$-gon and necklace are stationary non-rotating configurations in the laboratory frame. In each case, in the appropriate rotating frame the net force on each vortex due to the system boundary and the other vortices vanishes.}
        \label{figure:Ngonschematic}
\end{figure}
\section{Point Vortices}
\label{sec:pvmodel}
\subsection{Background}
\paragraph{Chiral systems} Systems of vortices with only one sign of circulation have been the subject of intense study due to their importance in superconductors and superfluid ground states. The most well-known stationary configuration of same-sign vortices is the Abrikosov lattice, a rigidly rotating periodic array with six-fold rotational symmetry that forms the energetic ground state in a rotating frame. Excited states in the rotating frame are chiral $N$-gon states, shown in a bounded domain in Figure~\ref{figure:Ngonschematic}. While the $N+1$-gon for $N=6$ has the same symmetry as the Abrikosov lattice, it is only a particular case of a general class of stationary solutions that are not energetic ground states, but rather form metastable equilibrium solutions. General existence proofs for periodic equilibria in bounded domains have been challenging, emerging only recently~\cite{kurakin_stability_2005,kadtke_method_1987,bartsch_critical_2015,bartsch_periodic_2016}.

\paragraph{Neutral systems} Systems with an equal number of positive and negative circulating vortices are much less studied, and play a central role in two-dimensional quantum turbulence where stirring a superfluid naturally injects vortex dipoles. Kuhl provides a general method for constructing vortex equilibria in any bounded, multiply connected domain with $n$-fold rotational symmetry~\cite{kuhl_equilibria_2016}. However, while quite general, the construction is rather opaque. Confining our attention to a disc-shaped domain, we examine neutral systems with even vortex number $N$. For a neutral $N =2$ system confined to a disc of radius $R$, the maximum energy state is also the stationary state, with the vortices lying along a common axis, at $x_\pm=\pm R/(2+\sqrt{5})^{1/2}\approx 0.486R$; this case has been studied theoretically~\cite{navarro_dynamics_2013,crasovan_globally_2002,crasovan_stable_2003,cawte_michael_m_snells_2019}, and experimentally~\cite{neely_observation_2010,middelkamp_guiding-center_2011}. For $N\geq 4$, the stationary state is no longer the maximum energy vortex configuration. A particular stationary state may be constructed by placing each vortex symmetrically in a ring around the center of the trap with alternating circulations, creating a neutral vortex necklace. For $N\geq4$ vortex necklaces have been studied theoretically in parabolic traps \cite{mottonen_stationary_2005,middelkamp_bifurcations_2010,theocharis_ring_2003,crasovan_globally_2002,crasovan_stable_2003}, and a constructive method of solution for arbitrary $N$ in this common BEC confinement is provided by Barry \emph{et al}~\cite{barry_generating_2015}. In the present work, we take an alternate approach to finding exact stationary solutions to the PVM for necklace $N$-gon states in a disc geometry with a hard wall boundary.

\subsection{Point-Vortex Model}
The equations of motion for $N$ point vortices in the plane are given by~\cite{helmholtz_lxiii._1867}
\begin{equation}
  {\dot{x}_i\choose\dot{y}_i} = \frac{\rho}{2\pi} \sum_{j=1}^N{\vphantom{\sum}}' \frac{\Gamma_j}{r_{ij}^2} {-y_{ij}\choose x_{ij}},
\end{equation}
where $\rho$ is the 2D fluid particle density, the prime on the summation denotes $j\neq i$, $\Gamma_j = \kappa_j h/m $ is the circulation, $\kappa_j = (-1)^j$ provides the sign of circulation, $x_{ij} = x_i - x_j$, and $r_{ij}^2 = (x_i-x_j)^2 + (y_i-y_j)^2$. As shown by Kirchhoff~\cite{kirchhoff_gustav_robert_vorlesungenuber_1876}, these equations of motion can be restated in Hamiltonian form
\begin{equation}
  \rho \Gamma_i \dot{x}_i = \frac{\partial H}{\partial y}, ~~~~~~~  \rho  \Gamma_i \dot{y}_i = -\frac{\partial H}{\partial x},
\end{equation}
where $x_i$ are the generalized coordinates and $p_i = \Gamma_i y_i$ are the generalized momenta conjugate to $x_i$. Converting to complex coordinates with $z_j = x_j + iy_j$ and subtracting off the infinite self-energy term gives the interaction Hamiltonian
\begin{equation}
  H(z_1,....,z_N) = -\frac{\rho}{2 \pi}   \sum_{i=1}^{N-1} \sum_{j=i+1}^{N} \Gamma_i \Gamma_j \log{ | z_{i} -  z_{j} |  }.
\label{eqn:1}
\end{equation}
In addition to the energy, in the disc domain the point-vortex angular momentum
\begin{align}\label{eq:conserved}
   I &= \sum_i \rho  \Gamma_i r_i^2,
\end{align}
is also conserved.

Lin~\cite{lin_motion_1941,lin_motion_1941-1} extended the dynamical equations to include bounded, simply-connected domains. For a disc of radius $R$, the method of images may be applied where each vortex has a corresponding image vortex with opposite circulation outside the disc at position $z_i^{\textrm{im}} = R /z_i^*$, where $z_j^* = x_j -iy_j$. The Hamiltonian becomes
\begin{equation}
   \label{eqn:hamiltonian}
   H(z_1,....,z_N) = -\frac{\rho }{2\pi}  \Bigg[\sum_{i=1,
   j\neq i}^N \Gamma_{i}\Gamma_{j} \log \frac{R| z_{i} - z_j |}{|R^2 - z_{i} z_{j}^* |}  - \frac{1}{2} \sum_{j}^N \Gamma_{j}^2 \log(R^2 - |z_{j} |^2)  \Bigg],
\end{equation}
and the equation of motion for vortex $i$ is
\begin{equation}
  \dot{z}_{i}^* = \frac{\rho}{2 \pi i}  \Bigg[  \sum_{j =1, 
  i\neq j}^N \frac{\Gamma_{j}}{ z_{i} -  z_{j} }  - \sum_{j=1}^N \frac{{\Gamma_{j} z_{j}^*}}{R^2 - z_{j}^* z_{i}}    \Bigg].
\label{eqn:PVmotion}
\end{equation}
The boundary terms allow for interesting stationary configurations involving a balance of intervortex forces and boundary forces. 

\section{Necklace N-gon: exact solution}
\label{sec:metalic_necklace}
We seek the $N$-fold rotationally symmetric stationary solution for $N$ point-vortex dipoles in the disc. We denote the position of each vortex in the disc as
\eqa{
z_n = d_N e^{i 2\pi (n-1)/N}
}
where $n= 1,2,3...$ is the vortex label, $d_N$ is the radial distance from the origin, and $N$ is the total number of vortices which is always an even integer; the precise dependence of $d_N$ on $N$ will be found. Substituting the vortex locations and the alternating circulation into the equation of motion (\ref{eqn:PVmotion}) for the first vortex gives
\eqa{
\dot{z}_{1}^* = -i\rho\frac{\Gamma}{2\pi} \left[ \sum_{n = 2}^{N} \frac{\kappa_n}{z_1 - z_n} - \sum_{n=1}^{N} \frac{\kappa_n z_n^*}{R^2 - z_n^* z_1 } \right],
}
The stationary states have $\dot{z}_{n}^*=0$ for all vortices, giving the equation
\eqa{
0 =   \sum_{n = 2}^{N} \frac{\kappa_n}{z_1 - z_n} - \sum_{n=1}^{N} \frac{\kappa_n z_n^*}{R^2 - z_n^* z_1 },
}
Due to the symmetry of the necklace and the disc geometry we only require the stationary solution for one vortex.
Since the first vortex sits on the positive $x$-axis, $z_1=d_N$, the steady state equation can be written as
\eqa{\label{Nvortexmotion}
S_1+S_2=0,
}
where the two sums
\begin{eqnarray}\label{s1}
S_1 &\equiv&    \sum_{n = 2}^{N} \frac{\kappa_n}{1 -  e^{i 2\pi (n-1)/N}}=\frac{1}{2},\\
S_2 &\equiv& - \left(\frac{d_N}{R}\right)^2\sum_{n=1}^{N}  \frac{\kappa_n  e^{-i 2\pi (n-1)/N}}{1-   (d_N/R)^2 e^{-i 2\pi (n-1)/N}} = \frac{N(R/d_N)^N}{1-(R/d_N)^{2N}},\label{s2}
\end{eqnarray}
are due to intervortex and image vortices respectively. Both sums are non-trivial to perform; the proofs are given in Appendix \ref{chap:S1}.
Using \eref{s1}, \eref{s2}, Eq.~\eref{Nvortexmotion} becomes
\eqa{\label{droots}
x^2-2Nx-1=0,
}
where $x=(R/d_N)^N$. The positive root yields $x=N+\sqrt{N^2+1}$, and the solution 
\eqa{\label{vloc}
d_N&=&\frac{R}{\left[N+\sqrt{N^2+1}\right]^{1/N}},
}
for the radial distance from the origin of the vortex necklace (the negative root locates the ring of image vortices~\footnote{In preparing this manuscript we discovered that this analytical result was found using a different approach in the doctoral thesis of Q. Dai \cite{bartsch_periodic_2016}.}. 
The solution uniquely determines a class of $N$-fold rotationally symmetric vortex necklace states as stationary solutions of the point vortex equations of motion in a disc domain. 
\begin{figure}[t!]
  \centering
       \includegraphics[width=0.6\columnwidth]{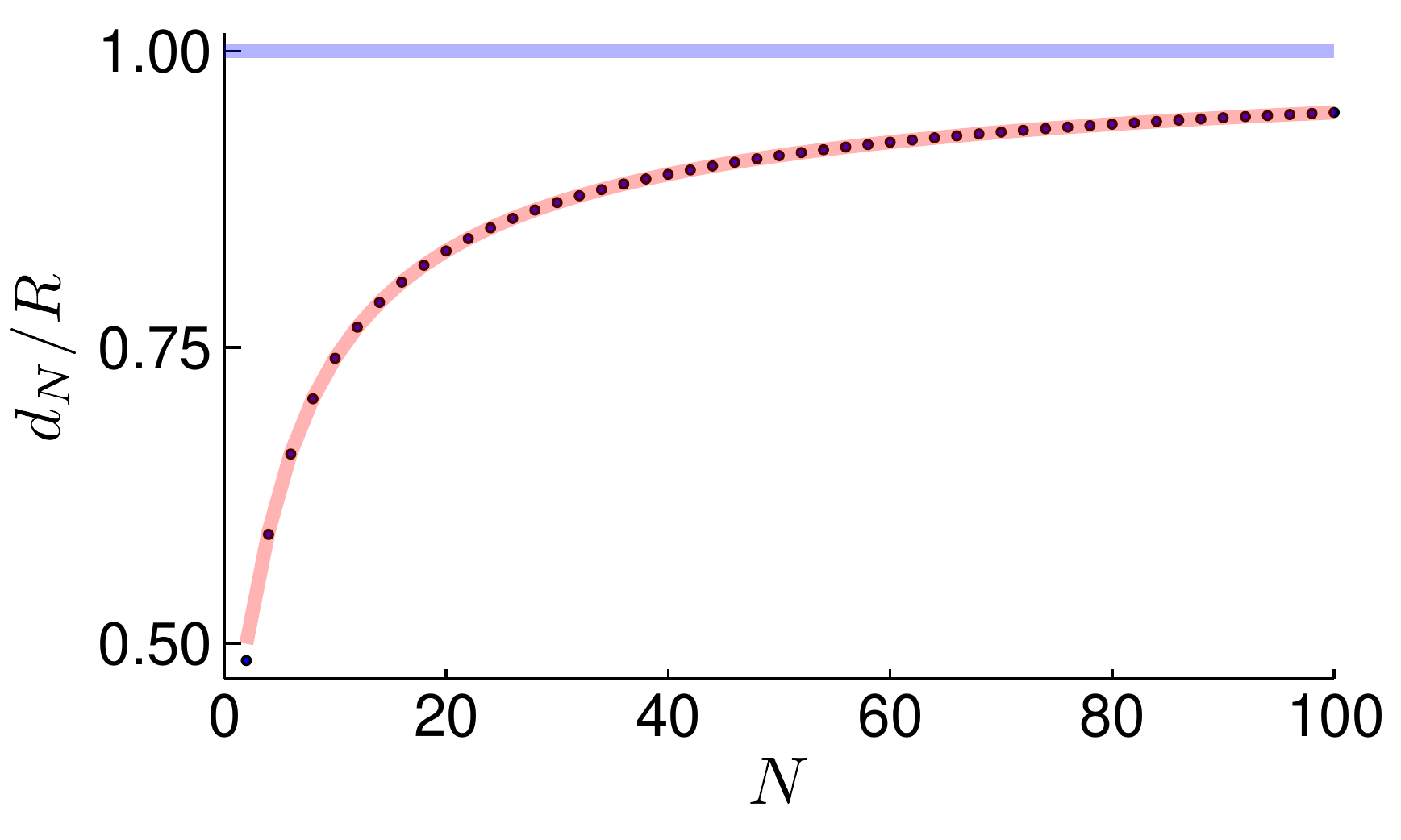}
    \caption{Vortex necklace radius for increasing $N$. Also shown (red line) is the solution for large $N$ given by $R/(2N)^{1/N}$, which is already a very good approximation for $N=4$.}
        \label{figure:necklaceRadius}
\end{figure}
We note that it is also possible two write this exact solution in terms of a generalization of the golden ratio $\varphi$, given by the positive root of the quadratic equation $
	\varphi^2 - \varphi -1 = 0$, namely $\varphi \equiv (1+\sqrt{5})/2$. The quadratic equation can be used to develop a continued fraction for $\varphi$:
\begin{equation}
	\varphi = 1 + \frac{1}{1 + \frac{1}{1 + \frac{1}{1 + \dots}}}.
\end{equation}
A generalization of the golden ratio known as the metallic mean was introduced by de Spinadel~\cite{de_spinadel_metallic_1999}
\eqa{\label{metm}
\Phi_n = n + \frac{1}{n + \frac{1}{n + \frac{1}{n + \dots}}} = \frac{n + \sqrt{n^2 + 4}}{2},
}
where $n \in \mathbb{N}$ and the $n$-th metallic mean, $\Phi_n$, is the positive root of $\Phi_n^2-n \Phi_n - 1 = 0$.
The first few metallic means are referred to as the identity (0-th), the golden (first), silver (second), and bronze (third) ratio. 

We note a useful approximation to the exact solution that will almost always be applicable. For a single dipole, $N=2$, and $d_2 = R/(2+\sqrt{5})^{1/2}\approx 0.4859 R$. However, for large $N\gg 1$,
\eqa{\label{dnapprox}
d_N \simeq& \frac{R}{(2N)^{1/N}}\left[1+\mathcal{O}\left(\frac{1}{N}\right)\right],
}
moving ever closer to the domain boundary at $R$ as $N$ increases. This behavior is shown in Figure~\ref{figure:necklaceRadius}, where it may also be seen that the approach to  $d_N\to R/(2N)^{1/N}$ is very fast, with an error of only 0.5\% for $N = 4$.
From Eq.~\eref{droots} we see that in terms of the metallic means the necklace states can be written as $d_N = R/\sqrt[N]{\Phi_{2N}}$; for large $N\gg 1$ \eref{metm} shows that  $\Phi_{2N}\to 2N$, consistent with \eref{dnapprox}. Already for $N=4$, $\Phi_{2N}\simeq 8.123\approx 2N$, with an error of only $\sim 1.5\%$. Geometrically, the position of each vortex is determined by $2N-1$ distances between the vortex, the $N-1$ other vortices, and the $N$ images imposed by the boundary. For $N\gg 1$ and $2N-1 \approx 2N$, and the self-interaction is unimportant relative to all of the mutual interactions between vortices and images. 

\section{\label{sec:simulation}Simulations}
The predictions of the PVM provide an approximation to the motion of vortices in a planar BEC confined by a hard-wall potential to a disc domain~\cite{gauthier_giant_2019,johnstone_evolution_2019}. In this section, we compare the exact PVM solutions with vortex dynamics governed by the Gross-Pitaevskii equation. We model a system of atoms held in a hard-wall trap that is otherwise homogeneous. The atomic density is thus homogeneous away from a sharp boundary, and the primary physics beyond the PVM stems from the compressibility of the quantum fluid: there is a small but non-negligible vortex core size set by the healing length. Closely related is the finite thickness of the boundary layer of the fluid. In general vortices can also couple to the phonon field supported by the BEC. Provided the healing length is small, as is the case in 2DQT systems, we may expect PVM and GPE to be in fairly close agreement~\cite{gauthier_giant_2019}. However, it is not a priori clear how close the agreement will be, nor the nature of the departure. We examine the dynamics of the vortex necklace $N$-gon, both for initial conditions very near the point vortex stationary solution and for larger deformations of the vortex configuration.

\subsection{Gross-Pitaevskii equation}
For the superfluid simulations we consider a BEC state described by wavefunction $\Psi(\mathbf{r},t)$, and containing $N$ atoms
\begin{equation}
\int d^3\mathbf{r}\;|\Psi(\mathbf{r},t)|^2=N.
\end{equation}

Numerically we solve the GPE~\cite{pitaevskii_bose-einstein_2003}
\begin{equation}
i\hbar\frac{\partial \Psi(\mathbf{r},t)}{\partial t} = \bigg( -\frac{ \hbar^2}{2m} \nabla^2 + V_{\rm ext}(\mathbf{r},t) + \varg|\Psi(\mathbf{r},t)|^2 \bigg)\Psi(\mathbf{r},t),
\end{equation}
where $V_{\rm ext} $ is the external potential, $\varg = 4\pi\hbar^2 a_s/m$ is the two-body interaction parameter, and $a_s$ is the s-wave scattering length. We assume the system is tightly confined in the $z$ direction so that the wavefunction is separable $\Psi(\mathbf{r},t)\equiv\psi(x,y,t)\psi_0(z)$. Integrating over the $z$ direction, the GPE for $\psi$ assumes a 2D form, but with interaction parameter $\g2d = \varg\int |\psi_0|^4 dz$; here the ground state $z$ wavefunction is a Gaussian normalized as $\int d^2\mathbf{r}\;|\psi_0|^2=1$. Defining the healing length
\begin{equation}
\xi \equiv \frac{\hbar}{\sqrt{m\mu}},
\end{equation}
where the chemical potential $\mu=\g2dn_0$ is determined by $n_0$, the homogeneous 2D atomic number density away from the steep $x-y$ confining potential  trapping potential.
For the homogeneous system, it is convenient to work in units of length, time, and energy of $\xi$, $\xi/c$, and $\mu=\hbar^2/m\xi^2$ respectively, where $c\equiv\sqrt{\g2dn_0/m}$ is the speed of sound.

\subsection{Initial conditions}
The disc-shaped hard-wall potential was constructed as
\begin{equation}
    V_{\rm ext}(x,y) =
\begin{cases}
     0,  &\for\;\sqrt{x^2 + y^2} < R,\\
     ~ 5\mu,               &\for\; \sqrt{x^2 + y^2} \geq R,
\end{cases}
\end{equation}
where $R$ is the radius of the trap. The total dimensions of the system were $128\xi \times 128\xi$ with $512$ grid points in each direction chosen to give high resolution inside the vortex cores. Each simulation was started with a fixed number of particles in the external potential $V_{\rm  ext}$ and evolved with imaginary time in order to find the ground state of the system. The wavefunction ansatz for the system is given by
\begin{equation}
\psi(\mathbf{r})= \prod_{n=1}^N f(\mathbf{r}-\mathbf{r}_n) e^{i\varphi (\mathbf{r})},
\label{eqn:phaseN}
\end{equation}
\begin{figure*}[t!]
\centering
\includegraphics[width=\columnwidth]{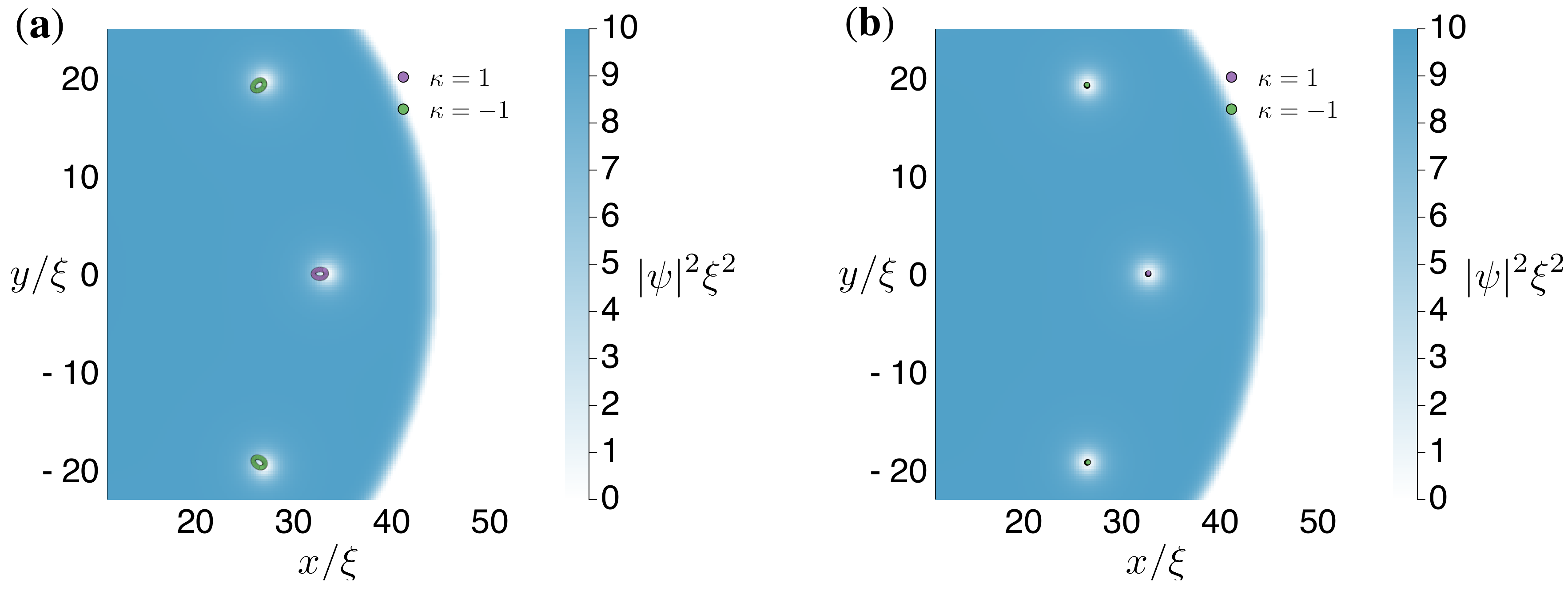}
\caption{GPE simulations of the $N=10$ necklace state in a disc of radius $R=45\xi$. The vortex trajectories are presented by a scatter plot superimposed on a heatmap of the initial particle density. \textbf{(a)}: GPE simulation of the $N=10$ necklace state after simulation duration $t = 2,000 \xi/c$. Each vortex in the necklace state orbits about a point on a length scale of about a healing length. \textbf{(b)}: GPE simulation of the $N=10$ necklace state radially perturbed inwards by $0.6\xi$ after duration $t = 2,000 \xi/c$. 
}
\label{figure:N10_stat}
\end{figure*}
where $f(\mathbf{r})$ is the profile of a single vortex in a homogeneous background condensate computed numerically to high accuracy using a Chebyshev basis~\cite{reeves_quantum_2017}, $\mathbf{r}_n$ is the location of each vortex, and $\varphi (\mathbf{r})$ is the phase. For a necklace state with well separated vortices, the solutions for the density profile of a single vortex may be multiplied together to create the density profile of the necklace state that gives a good approximation to the true solution. The location of each vortex in the necklace is given by
\eqa{
\mathbf{r}_n = (x_n,y_n)\equiv d_N [\cos{(2\pi (n-1)/N)},\sin{(2\pi (n-1)/N)}],
}
and the phase of the necklace state is given by
\begin{equation}
	\begin{split}
	\varphi(x,y) = \sum_{n=1}^N e^{i\pi(n-1)} \Bigg[ & \arctan\left(\frac{y-y_n}{x-x_n}\right) - 
	 \arctan\left(\frac{y-y_n^i}{x-x_n^i}\right)\Bigg],
	\end{split}
\end{equation}
where $y^i_n \equiv y_n R^2/|d_N|^2$ and $x^i_n \equiv x_n R^2/|d_N|^2$ are the positions of the image vortices which enforce the condition that fluid cannot flow across the boundary. These functions are used to imprint phase profile and the vortex density onto the ground state condensate at each vortex location. The use of the exact density profile of the single vortex core means that there is almost no additional acoustic energy introduced during the vortex creation process. 

\subsection{Corresponding units of length, time, and energy  between PVM and GPE simulations}

Units of length, time, and energy may be chosen for the PVM simulations such that the vortex dynamics are comparable to the GPE simulations. This can be done by choosing the unit of length to be $\xi$, the unit of time to be $2\pi \xi^2/\Gamma$, and the unit of energy to be $\rho\Gamma^2 /2\pi$ where $\rho$ is given by $\rho = m/2\pi \xi^2$. 

Although this choice of units for the PVM gives vortex dynamics comparable to the GPE system with identical initial vortex positions, the fluid that the GPE describes is compressible and includes finite vortex core size and boundary effects. Hence, the dynamics in each model for simulations with the same duration of time should not be expected to be identical.

\subsection{Computational algorithms}

The GPE and PVM were numerically integrated in the Julia programming language~\cite{bezanson_julia_2017} with the \texttt{Tsit5}~\cite{tsitouras_rungekutta_2011} algorithm from the  \texttt{DifferentialEquations.jl} package~\cite{rackauckas_differentialequations.jl_2017}. The GPE was integrated using a  pseudo-spectral Fourier method and vortices were created and detected using the \texttt{VortexDistributions.jl} package~\cite{bradley_a_s_vortexdistributions.jl_2019}.

\subsection{Stationary and Quasi-Stationary States}

The analytic solution to the necklace state was confirmed to be stationary in numerical PVM simulations. However, using the initial necklace state positions from the PVM in the GPE results in vortices with small periodic trajectories. As an example, we use the $N=10$ necklace state whose initial vortex positions in the PVM are given by $d_{10}$ (Eq. \ref{vloc}). A GPE simulation of this state is shown in Figure \ref{figure:N10_stat} $\textbf{(a)}$, in which the vortex trajectories precess inwards by $1.4\xi$ for many orbits. A correction to the PVM prediction for the necklace state is required as the condensate density must smoothly transition from the homogeneous interior solution to zero outside the hard-wall potential, inducing a shift in the velocity field due to displaced image charges.

To numerically identify the boundary adjustment to the PVM solution, we performed GPE simulations for $N =[4,6,8,10]$ necklace states in the $R=45\xi$ trap, with radially symmetric perturbations to the initial positions in increments $0.1\xi$ over the interval $-1\xi$ to $1\xi$. As shown in Appendix \ref{stationary}, a small adjustment to the PVM solution (\ref{vloc}) of less than $\xi$ is sufficient to locate the GPE stationary solutions. We refer to this small perturbation as the boundary correction, and in the case of the $N=10$ necklace state in Figure \ref{figure:N10_stat} $\textbf{(b)}$ we apply the boundary correction by radially perturbing the initial vortex positions inwards by $0.6\xi$. After a simulation of duration equal to that in Figure \ref{figure:N10_stat} $\textbf{(a)}$, the well-chosen boundary correction is seen to reduce the precession to be much less than $\xi$, imperceptible in the figure. 

\subsection{Angular momentum conserving deformations}
 
 \begin{figure}[t!]
 \centering
 \includegraphics[width=0.95\linewidth]{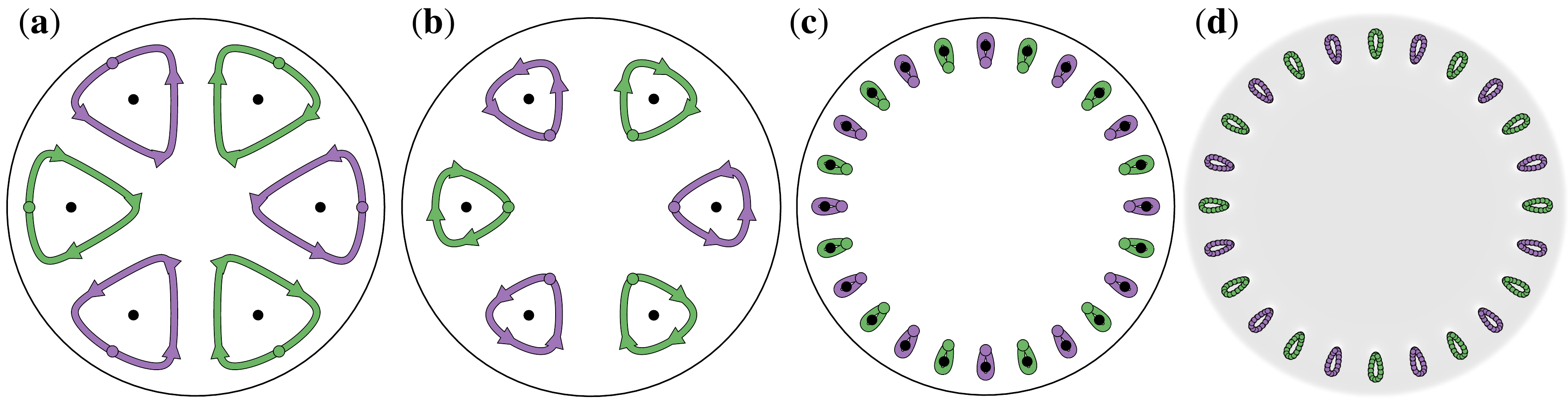}
 \caption{Large deformations to the vortex necklace that conserve angular momentum. The stationary positions are marked in black, the deformed position is marked with a coloured circle, and the arrows indicate the direction of motion. Each system is in a disc of radius $R = 45\xi$. \textbf{(a)}: PVM simulation for 1 period of oscillation ($t \approx 1200 \xi/c$) for the $N=6$ necklace radially deformed outwards by $10\xi$. \textbf{(b)}: PVM simulation for 1 period of oscillation ($t \approx 1519 \xi/c$) for the $N=6$ necklace state deformed radially inwards by $10\xi$. \textbf{(c)}: PVM simulation for 1 period of oscillation ($t = 360 \xi/c$) for the $N=24$ necklace state deformed inwards by $4\xi$. \textbf{(d)}: GPE simulation for 2.8 periods of oscillation for the $N=24$ necklace state deformed inwards by $4\xi$. The simulation had a duration of $t = 1,000 \xi/c$. Here the stationary positions marked in black are omitted for visual clarity and the lines are replaced by scattered points representing the location of the vortices through time.
 }
 \label{figure:Fig5}
 \end{figure}

 Non-perturbative, large-amplitude deformations of the vortex necklace states induce interesting quasi-periodic vortex dynamics. In the remainder of this work, we study two classes of deformations, comparing PVM and GPE simulations. In Figure \ref{figure:Fig5} (\textbf{a}) and \ref{figure:Fig5} (\textbf{b}), we show some simple deformations to the necklace state in the point-vortex model that preserve the angular momentum $I$, Eq.~(\ref{eq:conserved}). The vortex trajectories resulting from these two deformations trace out similar triangular orbits, with the outward deformation creating larger orbits.

We next construct the same kind of angular momentum conserving deformation to the $N=24$ necklace state by radially perturbing the necklace state inwards by $4\xi$, but this time compare the dynamics between PVM and GPE evolution. The PVM dynamics are presented in Figure \ref{figure:Fig5} (\textbf{c}) and the GPE dynamics in Figure \ref{figure:Fig5} (\textbf{d}). The vortex trajectories in both systems now trace out somewhat elliptical orbits, with a faster period of oscillation than the previous deformations. An outwards deformation of the $N=24$ necklace of magnitude $10\xi$ is unfeasible as the initial vortex positions would exit the system. An inward deformation of the same magnitude is unstable in GPE simulations, however in PVM simulations it results in vortex trajectories tracing out large semi-triangular orbits. 

\subsection{Non-conserving deformations: bow-star perturbation}
The vast majority of perturbations to the necklace state that do not conserve angular momentum produce chaotic trajectories. In this section we present a particular perturbation to the necklace state that produces periodic trajectories that are qualitatively distinct from the previous orbits. This new perturbation turns out to be more sensitive to differences between the PVM and GPE theories of vortex dynamics in the disc domain. We find that correcting for the exterior boundary thickness again gives better agreement between PVM and GPE. 

The perturbation of a necklace state of $N$ vortices is produced by moving all vortices of one circulation radially inwards by the constant $p_N e^{\pi /N}$, and all vortices of the opposite circulation radially outwards by the constant $p_Ne^{-\pi /N}$. With this choice of perturbation, the radially inwards perturbed vortices trace out a periodic star-shaped trajectory with the number of vertices equal to $N$, and the radially outward perturbed vortices trace out a trajectory that loops around each vertex of the star in turn. We refer to the perturbation, discovered numerically, as the \emph{bow-star perturbation:}
\begin{equation}
	z_n =  \sum_{n=1}^{N} \left[ d_N + p_N  e^{(-1)^{(n-1)} \pi/N}\right] e^{2\pi i(n-1)/N}.
	\label{eqn:bowstar6}
\end{equation}
Vortex trajectories resulting from the bow-star perturbation deviate from periodic behaviour into chaos with small changes to initial conditions. Empirical approximations to $p_N$ require many significant digits to produce the periodic bow-star trajectory. This bow-start perturbation may provide an interesting starting point for seeding vortex chaos.

\begin{figure}[t!]
  \centering
  \includegraphics[width=0.98\linewidth]{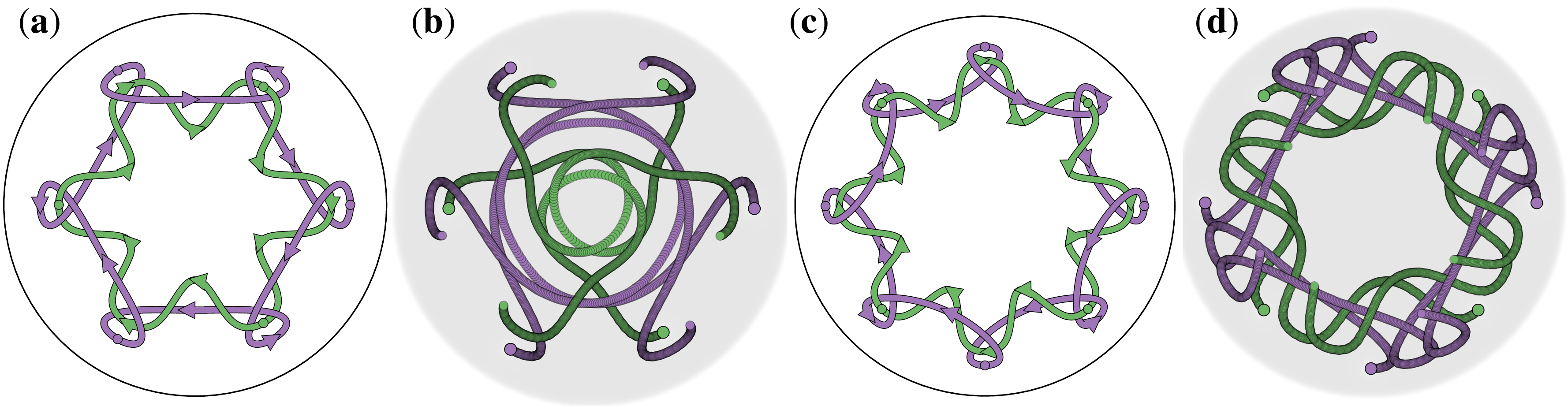}
  \caption{PVM and GPE simulations of $N=6$ and $N=8$ bow-star trajectories perturbed from the necklace state by Eq. \ref{eqn:bowstar6}. The lines in the PVM figures trace out the trajectories of the vortices through time with the arrows point in their direction of motion. The vortex trajectories in the GPE simulations are presented with scatter plots. The systems all have a disc of radius $R = 45 \xi$. \textbf{(a)}:PVM simulation of the $N=6$ bow-star state for 3 periods of oscillation in which each vortex returns to its own initial position. The simulation had a duration of $t = 34,740 \xi/c$. \textbf{(b)}: GPE simulation of duration $t = 3,260 \xi/c$ for the $N=6$ bow-star state. \textbf{(c)}: PVM simulation of the $N=8$ bow-star state for 2 periods of oscillation in which each vortex returns to its own initial position. The simulation had a duration of $t = 21,650 \xi/c$. \textbf{(d)}: GPE simulation of duration $t = 3,240 \xi/c$ for the $N=8$ bow-star state. }
  \label{figure:Fig7}
  \end{figure}
  
  An example of the bow-star perturbation in Eq. \ref{eqn:bowstar6} with $N=6$ is shown in Figure \ref{figure:Fig7}, where the value of $p_6=4.003038160663878...$ was found empirically. The PVM simulation in Figure \ref{figure:Fig7} (\textbf{a}) resulting from this perturbation produces a regular motion in which each vortex returns to its own initial position of period $t \approx 11,580 \xi/c$. The negatively charged vortices trace out a six-pointed star and the positively charged vortices create a bow around the vertices of the star. The GPE simulation in Figure \ref{figure:Fig7} (\textbf{b}) traces out a shape with the same "star" points and "bows", but each vortex rotates around the origin before tracing out the next star point or bow. Here, we only evolve the GPE simulation long enough to see the trajectories diverging from the PVM.

  By creating the $N=8$ necklace state with same perturbation from Eq. \ref{eqn:bowstar6} but with $p_8=4.022725...$. The PVM simulation of the $N=8$ bow-star in Figure \ref{figure:Fig7} (\textbf{c}) has a period of $t \approx 10,825 \xi/c$ whereby each vortex returns to its own initial position. In these dynamics, negatively charged vortices trace out an eight-pointed star and the positively charged vortices create a bow around each star point. In the GPE simulation of the $N=8$ bow-star in Figure \ref{figure:Fig7} (\textbf{d}), the negatively charged vortices trace out a shape close to a four-pointed star, with the bows from the positively charged vortices now in-between the stars. The shape becomes rotationally shifted before the first period.

  \begin{figure}[t!]
    \centering
    \includegraphics[width=0.95\linewidth]{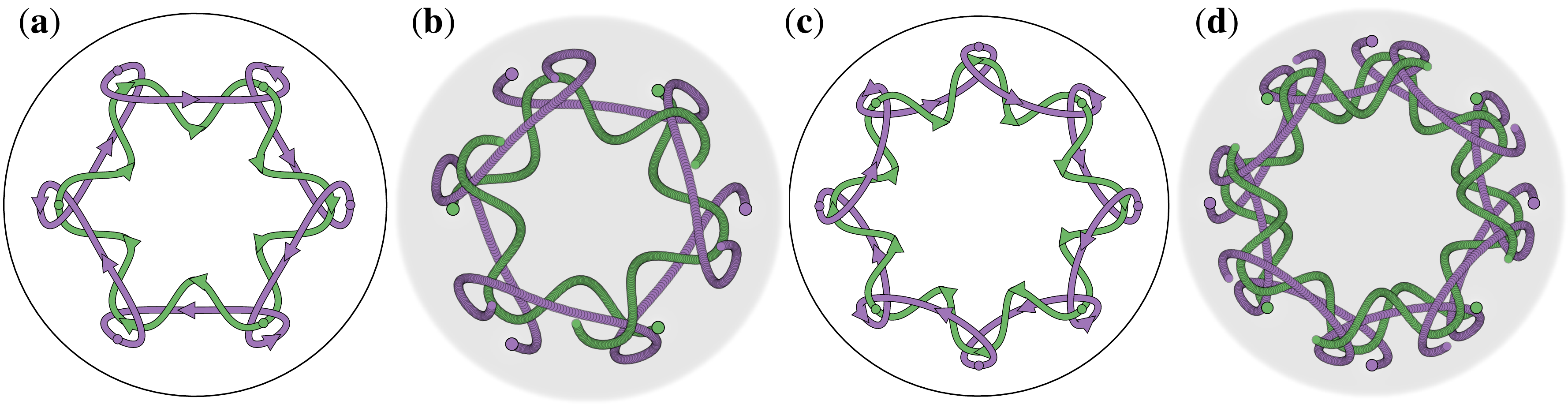}
    \caption{$N=6$ and $N=8$ necklace states with the bow-star perturbation given in Eq. \ref{eqn:bowstar6} and an inwards radial perturbation of $0.6\xi$ applied. The lines in the PVM figures trace out the trajectories with arrows indicating direction. The vortex trajectories in the GPE simulations are presented with scatter plots (gray background). The systems all have a disc of radius $R = 45 \xi$. \textbf{(a)}: PVM simulation of the $N=6$ bow-star state for 3 periods of oscillation in which each vortex returns to its own initial position. The simulation had a duration of $t = 34,743 \xi/c$. \textbf{(b)}: GPE simulation of the boundary corrected $N=6$ bow-star state of duration $t = 4,280 \xi/c$. \textbf{(c)}:  PVM simulation of the $N=8$ bow-star state for 3=2 periods of oscillation in which each vortex returns to its own initial position. The simulation had a duration of $t = 21,650 \xi/c$. \textbf{(d)}: GPE simulation of the boundary corrected $N=8$ bow-star state of duration $t = 4,045 \xi/c$.}
    \label{figure:Fig9}
    \end{figure}
    As demonstrated in Figure \ref{figure:N10_stat} \textbf{(b)}, the necklace state in the GPE does not agree with the predictions of the necklace state from the PVM. By scanning over the parameter space a suitable correction to the necklace state in the GPE was found in Appendix \ref{sec:Appendix2}. Although the bow-star trajectories are different between the PVM and GPE simulations, we may adjust the initial bow-star positions in the GPE by including the boundary correction
\begin{equation}
	z_n =  \sum_{n=1}^{N} \left[ d_N -0.6 + p_N  e^{(-1)^{(n-1)} \pi/N}\right] e^{2\pi i(n-1)/N}.
	\label{eqn:adjustedbowstar}
\end{equation}
The results of this adjustment to the initial positions of the bow-star state are shown in Figure \ref{figure:Fig9}. The $N=6$ bow-star with the boundary correction is shown in Figure \ref{figure:Fig9} (\textbf{a}) after GPE evolution of duration $t = 4,280 \xi/c$. The trajectory is now much closer to that of the point-vortex trajectory in Figure \ref{figure:Fig7} (\textbf{a}), with the vortices tracing out an off-resonance 6 pointed bow-star shape. In Figure \ref{figure:Fig7} (\textbf{d}), we see the $N=8$ bow-star with corrected initial positions after GPE simulation of duration $t = 4,045 \xi/c$. The trajectory is now closer to that of the point-vortex trajectory in Figure \ref{figure:Fig7} (\textbf{c}), with the vortices tracing out an 8 pointed bow-star shape.

\subsection{Non-Neutral System}
We have focused on the neutral system, but we note that our approach may also be used to find the stationary $N+1$ vortex polygon, in which $N$ vortices of the same circulation are placed in the disc radially symmetrically around a central vortex of oppositely signed circulation and varying strength.

The circulation of the ring vortices is given by $\Gamma_r = \kappa_r h/m$ where $\kappa_r \in \mathbb{R}$ is the strength and sign of circulation, and similarly $\Gamma_c = \kappa_c h/m$ for the central vortex which must have opposite sign compared to the radial vortices. The radial distance for the $N$ vortices in the $N+1$ vortex polygon is given by
\begin{equation}
 d = R\left( \frac{2 \kappa_c +\kappa_r N -\kappa_r}{2\kappa_c - \kappa_r N -\kappa_r} \right)^{\frac{1}{2N}},
\end{equation}
provided $N\geq 2$ and the relation
\begin{equation}
\frac{\kappa_c}{\kappa_r} <  \frac{(1-N)}{2} 
	\label{eqn:npone}
\end{equation}
 is satisfied. In the PVM vortex strengths of any real number are permitted provided Eq. \ref{eqn:npone}  is satisfied. However, in a BEC vortices must have integer circulations and vortices with charge $\geq 2$ are unstable. For example, by selecting $N=2$, $\Gamma_r = -h/m $ for the radial circulations, and $\Gamma_c = h/m$ for the central vortex circulation, the conditions required for a BEC system are satisfied with the distance of the radial vortices being $d = \sqrt[5]{4}$ plus a boundary correction.

\section{Discussion and Conclusions}\label{sec:conclusions}

We have found an exact solution for the neutral vortex necklace in a planar superfluid confined to a disc shaped domain. The radial locations of each vortex, due to a balance of forces from other vortices and the outer fluid boundary, can be written in terms of the \textit{metallic mean}, a generalized golden ratio. We have tested the stationary point-vortex solutions in simulations of the Gross-Pitaevskii equation, finding that a small radial adjustment, of order a healing length, yields a good approximation to the metastable state for a dilute gas Bose-Einstein condensate. Comparing the dynamics of point vortex model and Gross-Pitaevskii equation, we find that large scale deformations generate interesting symmetric and quasi-periodic vortex trajectories. The link between the form of perturbations and the transition between ordered periodic trajectories and chaos remains an interesting open question. Our results indicate that neutral vortex necklace may be an accessible metastable non-rotating state in atomic Bose-Einstein condensates, given sufficient vortex control, and may offer an interesting route to seed chaotic and turbulent vortex dynamics. 

\section{Acknowledgements}
We are grateful to Xiaoquan Yu for stimulating discussions. This work was supported by Marsden Grant UOO1762.
\appendix

\section{\label{chap:S1}Vortex Sums}
\subsection*{\label{sec:Appendix1}Inter-Vortex Terms}

The sum $S_1$ gives the contributions to the motion of $z_1$ from each of $N-1$ vortices
\eqa{
S_1 =     \sum_{n = 2}^{N} \frac{e^{i\pi n}}{1 -  e^{i 2\pi (n-1)/N}}.
}
If the system being solved consists of a single dipole, then the total number of vortices is $N = 2$ and the summation $S_1$ is reduced to the single term
\eqa{
S_1 =   \left( \frac{e^{i\pi 2}}{1 -  e^{i 2\pi (2-1)/N}}\right) =  \frac{1}{2}.
}
As we now show, this result holds for any even number of vortices.  
If the number of dipoles is greater than one, then we can shift the index of the summation and use the fact that the sum is always over an odd number of terms to find
\eqa{
S_1 =  -  \sum_{k = 1}^{N-1} \frac{e^{i\pi k}}{1 -  e^{i 2\pi k/N}}=
  - \Bigg[  \sum_{k = 1}^{N/2-1} \frac{e^{i\pi k}}{1 -  e^{i 2\pi k/N}} +  \frac{e^{i\pi N/2}}{1 -  e^{i 2\pi (N/2)/N}}  
+ \sum_{k = N/2 + 1}^{N-1} \frac{e^{i\pi k}}{1 -  e^{i 2\pi k/N}} \Bigg],
}
or
\begin{equation}
  S_1 =   -\frac{e^{i\pi N/2}}{2} 
  -   \left[  \sum_{k = 1}^{N/2-1} \frac{e^{i\pi k}}{1 -  e^{i 2\pi k/N}}  + \sum_{k = N/2 + 1}^{N-1} \frac{e^{i\pi k}}{1 -  e^{i 2\pi k/N}} \right].
  \end{equation}
  The second sum can be cast in terms of the same index as the first. We first shift the index on the second sum
  \begin{equation}
  S_1 =   -\frac{e^{i\pi N/2}}{2} 
  -   \left[  \sum_{k = 1}^{N/2-1} \frac{e^{i\pi k}}{1 -  e^{i 2\pi k/N}}  + \sum_{k = 1}^{N/2-1} \frac{e^{i\pi (k+N/2)}}{1 -  e^{i 2\pi (k+N/2)/N}} \right],
  \end{equation}
  and then invert the order of summation by letting $l = N/2-k$ and substituting it into the second sum, and finally relabel $l\to k$ to give
  \begin{equation}
  S_1 =   -\frac{e^{i\pi N/2}}{2} 
  -    \sum_{k = 1}^{N/2-1} \left[ \frac{e^{i\pi k}}{1 -  e^{i 2\pi k/N}}  +  \frac{e^{i\pi (N-k)}}{1 -  e^{i 2\pi (N-k)/N}} \right].
  \end{equation}
  Using the fact that $e^{i\pi k}=e^{i\pi (N-k)}$ since $N$ is even, we have
  \begin{equation}
  S_1 =   -\frac{e^{i\pi N/2}}{2} -    \sum_{k = 1}^{N/2-1} e^{i\pi k} \left[ \frac{1}{1 -  e^{i 2\pi k/N}}  +  \frac{1}{1 -  e^{-i 2\pi k/N}} \right]
  = -\frac{e^{i\pi N/2}}{2} -    \sum_{k = 1}^{N/2-1} e^{i\pi k} 
  =   \frac{(-1)^{N/2-1}}{2} -    \sum_{k = 1}^{N/2-1} e^{i\pi k}\nonumber.
  \end{equation}
  If $N/2-1$ is even, the first term is $1/2$ and the second is $0$; if $N/2-1$ is odd the first term is $-1/2$ and the second is $1$. Hence $S_1= 1/2$.

  \subsection*{\label{sec:Appendix2}Vortex Image Terms}

The second summation is the contribution from all of the vortex-images to the motion of the vortex $z_1$. It is given by
\eqa{
S_2 = - \left(\frac{d_N}{R}\right)^2\sum_{n=1}^{N}  \frac{\kappa_n  e^{-i 2\pi (n-1)/N}}{1-   (d_N/R)^2 e^{-i 2\pi (n-1)/N}  }.
\label{eqn:a10}
}
To simplify, we let $a = (d_N/R)^2$
\begin{equation}
S_2= -a\sum_{n=1}^N\frac{e^{i\pi n} e^{-i(n-1)2\pi/N}}{1-a e^{-i(n-1)2\pi/N}} 
= a\sum_{k=0}^{N-1}\frac{e^{ik(N/2-1)2\pi/N}}{1-a e^{-ik2\pi/N}}.
\end{equation}
The geometric series Fourier transform pairs, valid for $a\neq 1$ are
\eqa{\label{fs1}
\sum_{j=0}^{N-1}e^{-ijk2\pi/N}\;a^j=\frac{1-a^N}{1-ae^{-ik2\pi/N}},
}
\eqa{\label{fs2}
\frac{1}{N}\sum_{k=0}^{N-1}e^{ijk2\pi/N}\frac{1-a^N}{1-ae^{-ik2\pi/N}}=a^j,
}
where $j$ can be chosen either positive or negative (the latter requiring $a\to a^{-1}$). Since we have initially made the choice $(-1)^k=e^{ik\pi}$, the transform \eref{fs2} with positive choice $j\equiv N/2-1$ gives
\eqa{
a\sum_{k=0}^{N-1}\frac{e^{ik(N/2-1)2\pi/N}}{1-ae^{-ik2\pi/N}}=aN\frac{a^{(N/2-1)}}{1-a^N},
\label{eqn:a14}
}
with the result
\eqa{\label{s2res1}
S_2=\frac{N(d_N/R)^N}{1-(d_N/R)^{2N}}.
}
If we combine \eref{s1} and \eref{s2res1}, Eq. (\ref{Nvortexmotion}) becomes
\begin{equation}
\frac{1}{2}+\frac{N(d_N/R)^N}{1-(d_N/R)^{2N}}=0,
\end{equation}
or, in terms of $x=(d_N/R)^N$, the equation $x^2-2Nx-1=0$, with positive root $x=N+\sqrt{N^2+1}$ giving
\eqa{
d_N/R=(N+\sqrt{N^2+1})^{1/N} >1,
}
that describes the location of the images. The vortex locations are given by the second allowed choice for the phases: $(-1)^k=e^{ik\pi}$, and $j=-N/2-1$, instead giving the sum \eref{s2} and the location of the vortices \eref{vloc}.

\section{GPE stationary states from point-vortex}\label{stationary}
\begin{figure*}[t!]
  \centering
    \includegraphics[width=0.8\textwidth]{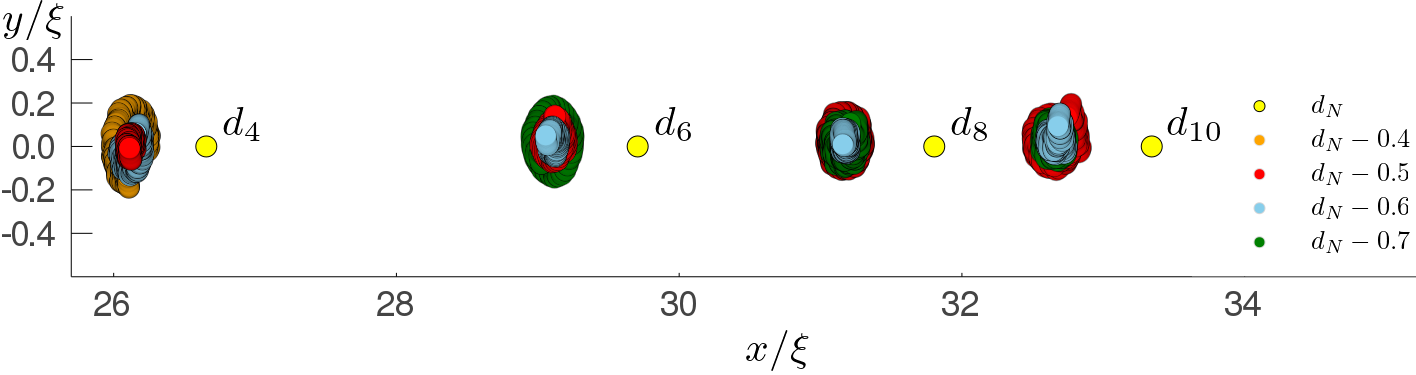}
    \caption{Trajectories from GPE simulations of perturbed necklace states. Only the trajectories of the first vortex in each state are shown. $d_4,d_6,d_8$ and $d_{10}$ mark the positions of the PVM stationary necklace state. To the left of each PVM stationary state marker are the trajectories of the first vortex in each state for different perturbed initial conditions.}
        \label{figure:all_stat}
\end{figure*}

In Figure \ref{figure:all_stat} the $N = 4,6,8,10$ stationary PVM necklace state positions of the first vortex in each necklace are marked in yellow. To the left of each PVM stationary position marked in yellow, are the perturbed GPE trajectories of the first vortex in each necklace simulated over duration $t = 2,000 \xi/c$. To the left of the $N=4$ PVM stationary position are three perturbed GPE trajectories. The trajectories that had their initial positions perturbed from the PVM stationary position by $-0.4$ and $-0.6$ are shown in orange and sky-blue respectively. These trajectories orbit a common center. The trajectory with initial position perturbed by $-0.5$ stays within a range of $0.1\xi$ from the duration of the simulation.

 The GPE trajectories of the perturbed $N=6$ necklace state are to the left of the $d_6$ yellow marker. The trajectory of the first vortex in the necklace perturbed inwards by $-0.6\xi$ from the PVM prediction is shown in sky-blue, and stays within a range of $0.1\xi$ over duration $t = 2,000 \xi/c$. Similarly, the perturbation of $-0.6\xi$ from the PVM leads to a tight oscillation around the true stationary point for both the $N=8$ and $N=10$ necklace respectively. For necklace states with $N>10$, the GPE trajectories become unstable on increasing shorter time scales.





%

\begin{acknowledgements}
MC and AB acknowledge financial support from the Marsden Fund (Grant No. UOO1726), and the Dodd-Walls Centre for Photonic and Quantum Technologies. 
MR was was partially supported by the Australian Research Council Centre of Excellence in Future Low-Energy Electronics Technologies (Project No.  CE170100039), and by the Australian government.
\end{acknowledgements}

%
%



\newcommand{\noopsort}[1]{}

\end{document}